\documentclass[aps,a4paper,pre,showpacs,10pt,floats,floatfix,nofootinbib,superscriptaddress,twocolumn]{revtex4}

\usepackage{amsmath}
\usepackage{amsfonts}
\usepackage{amssymb}
\usepackage[dvipdf]{graphicx}
\usepackage{subfigure}
\usepackage[pdfborder={0 0 0}]{hyperref}  
\usepackage{color}
\usepackage[latin1]{inputenc}
\usepackage{float}
\usepackage{comment}

\begin{document}
 \title{Quasi-stationary simulations of the contact process on quenched networks}
 
\begin{abstract}
We present high-accuracy quasi-stationary (QS) simulations of the
contact process in quenched networks, built using the configuration
model with both structural and natural cutoffs. The critical behavior
is analyzed in the framework of the anomalous finite size scaling
which was recently shown to hold for the contact process on annealed
networks. It turns out that the quenched topology does not
qualitatively change the critical behavior, leading only (as expected)
to a shift of the transition point. The anomalous finite size scaling
holds with exactly the same exponents of the annealed case, so that we
can conclude that heterogeneous mean-field theory works for the
contact process on quenched networks, at odds with previous claims.
Interestingly, topological correlations induced by the presence of the
natural cutoff do not alter the picture.
\end{abstract}

\author{Silvio C. Ferreira} \email{silviojr@ufv.br}\thanks{On leave at
Departament de F\'{\i}sica i Enginyeria Nuclear, Universitat Polit\`ecnica de
Catalunya, Barcelona, Spain.}

\author{Ronan S. Ferreira}
\affiliation{Departamento de F\'{\i}sica, Universidade Federal de
  Vi\c{c}osa, 36571-000, Vi\c{c}osa - MG, Brazil}

\author{Claudio Castellano} 
\affiliation{Istituto dei Sistemi Complessi (ISC-CNR), via dei Taurini 19,
I-00185 Roma, Italy}
\affiliation{Dipartimento di Fisica, ``Sapienza''
  Universit\`a di Roma, P.le A. Moro 2, I-00185 Roma, Italy}

\author{Romualdo Pastor-Satorras} 
\affiliation{Departament de
  F\'{\i}sica i Enginyeria Nuclear, Universitat Polit\`ecnica de
  Catalunya, Campus Nord B4, 08034 Barcelona, Spain}


\maketitle

\section{Introduction}
\label{sec:intro}

Dynamical processes are strongly affected by the structure of the
pattern which mediates interactions. This common wisdom, whose
evidence has been confirmed by decades of investigations in
statistical mechanics, has assumed a new relevance in recent years,
with the advent of complex networks theory. Percolation, epidemic
spreading or synchronization are among the phenomena for which the
effect of a disordered topology has been more thoroughly
studied~\cite{CallawayPRL2000, pv01a, ArenasPhysRep2008}, but for many
other types of dynamics a complex substrate structure also induces
novel and nontrivial dynamical
behavior~\cite{dorogovtsevRMP08,barratbook}. This occurs in particular
when the pattern of connections is extremely heterogeneous, as in
scale-free (SF) networks, where the probability that an element is
connected to $k$ others (the degree distribution) is given by a
power-law form $P(k)\sim k^{-\gamma}$ \cite{barabasi02}.

Along with the recognition that dynamics on networks can be very
different from lattices, comes the natural question about how
theoretical methods can be adapted to deal with topologically complex
substrates. It has been early recognized that the standard mean-field
approach must be modified on strongly heterogeneous networks, to take
into account the broad variability in the connectivity of
vertices~\cite{pv01a}. Thus, in the heterogeneous mean-field (HMF)
theory~\cite{dorogovtsevRMP08,barratbook} the order parameter is
replaced by a set of analogous quantities which depend explicitly on
the degree $k$ of the node considered (the degree of a node is the
number of other nodes to which it is directly linked). This simple
modification turns out to describe with remarkable accuracy the
behavior of many systems, reinforcing the na\"{\i}ve expectation that
mean-field methods must work on networks, due to their
infinite-dimensional nature.

For this reason it came as a surprise when the numerical investigation
of the behavior of the contact process (CP) on generic power-law
distributed networks gave results in apparent disagreement with the
predictions of HMF theory~\cite{RomuPRL2006}. The contact
process~\cite{harris74} is an extremely simple model for spreading,
which, in a general network, is defined as follows: Vertices can be in
two different states, either empty or occupied. The dynamics includes
the spontaneous annihilation of occupied vertices, which become empty
at unitary rate, and the occupation of empty neighbors by occupied
vertices, with rate $\lambda/k_i$, where $k_i$ is the degree of the
occupied node. The model is characterized by a phase transition at a
value of the control parameter $\lambda = \lambda_c$, separating an
active phase from an absorbing one, devoid of occupied
vertices~\cite{marro1999npt}.

In Ref.~\cite{RomuPRL2006} the HMF theory for CP was derived in the
limit of infinite network size. Its predictions could not be directly
checked against numerical simulations because of the presence of
extremely large finite size effects. A comparison was made possible by
the introduction of a finite size scaling (FSS) ansatz~\cite{cardy88},
adapted to a network topology, leading to the conclusion that CP
dynamics on quenched networks was not described by the HMF
approximation.

Such a claim was criticized by Park and collaborators~\cite{HaPRL2007,
RomuPRL2007}, which later proposed an alternative FSS ansatz, based on
a droplet excitation theory~\cite{HongPRL2007}. A discrimination
between the two approaches turned out to be nontrivial: Even for
simulations on annealed networks, which are expected to be described
exactly by mean-field theory, numerical results did not satisfactorily
conform to any of the two competing theories~\cite{RomuPRL2007}.

Clarifying results on this issue have been obtained
later~\cite{RomuPRL2008,bogunaPRE2009}, revealing that FSS on annealed
networks is actually more complicated than previously assumed. At odds
with what happens in lattices, the behavior of the CP on networks
of finite size depends not only on the number of vertices $N$ but
also on the moments of the degree distribution. This last feature
implies, for SF networks, that the scaling around the
transition depends explicitly on how the largest degree $k_c$ diverges
with $N$. Such a dependence (which both previous FSS approaches were
lacking) introduces very strong corrections to scaling. However, if
such corrections are properly taken into account, it is possible to
show that the CP on annealed networks agrees, with high accuracy, with
the predictions of HMF theory~\cite{Ferreira_annealed}. A detailed
comparison between HMF and precise numerical results for annealed
networks has been  possible by using the quasi-stationary (QS)
state simulation method~\cite{Mancebo2005,DickmanJPA}. This is an
optimized numerical technique which prevents the system from falling
into the absorbing configuration, thus allowing the detailed
investigation of the order parameter close to the transition for
finite systems.

In the present paper, we go beyond the results of
Ref.~\cite{Ferreira_annealed} and exploit QS simulations for studying
the CP transition on quenched SF networks. We assume that finite size
scaling on quenched networks has exactly the same anomalous form valid
on annealed ones. In this way we are able to determine, with an
unprecedented accuracy, the location of the critical point and the
value of the exponents associated with the transition. It turns out
that the exponents computed numerically for quenched networks are in
excellent agreement with the HMF predictions: The CP on networks
definitely obeys heterogeneous mean-field theory. This result fills
the gap in the debate about the validity of HMF theory for CP. Both
competing theories~\cite{RomuPRL2006,HongPRL2007} were not correct.
The apparent discrepancy between theory and numerics reported in
Ref.~\cite{RomuPRL2006} was due to an incorrect FSS theory and the
presence of large corrections to scaling in simulations. The
alternative FSS approach of Ref.~\cite{HongPRL2007} was incorrect as
well. The correct FSS theory is more complicated than previously
assumed, but perfectly accounts for numerical results on both annealed
and quenched networks.

The present paper is structured as follows: In Section II we describe
the quenched networks considered, the contact process dynamics, and
the QS simulation method. The numerical determination of the
transition point and the associated exponents are presented in Sec.
III. Our concluding remarks are discussed in Sec. IV.

\section{Model and simulation methods}
\label{sec:model}

The CP is simulated on quenched networks with $N$ vertices and degree
sequence $\lbrace k_1,k_2,\ldots,k_N\rbrace$. The networks are built
according to the configuration model (CM)~\cite{molloy95}, where the
degree of each vertex is defined at the beginning of the simulation as
a random variable with a power law distribution $P(k)=Ak^{-\gamma}$
and $k_0 \leq k \leq k_c$. The degree distribution has a hard degree
cutoff $k_c=N^{1/\omega}$ (degrees larger than $k_c$ are not allowed),
where $\omega \ge \gamma-1$ is the cutoff exponent. This choice
reduces sample to sample fluctuations that make the critical analysis
difficult even for the simpler case of annealed networks
\cite{RomuPRL2008,bogunaPRE2009,NohPRE2009}.

Two models for the link assignment are investigated, namely, random
and ordered configuration models. In the random CM \cite{Catanzaro05},
two stubs (not connected links) are selected at random and the
respective nodes connected, avoiding self and multiple connections.
The procedure is repeated until all stubs are joined. In the ordered
CM nodes are sorted in descending degree order. Then a stub of the
most connected node is connected to another randomly selected stub,
avoiding self and multiple connections. After all stubs of the largest
vertex are connected, the next most connected node with free stubs is
picked up and the procedure is repeated until all stubs are connected.
The ordering of nodes allows to build large networks even for
$\gamma-1 \le\omega<2$, which is usually impossible with random
selection of nodes. A consequence of this nonrandom procedure is that
disassortative degree correlations are
generated~\cite{PhysRevLett.89.208701}.

The CP simulations on an arbitrary network are performed with the
standard protocol~\cite{marro1999npt}: At each time step, an occupied
vertex $j$ is chosen at random and time is updated as $t\rightarrow
t+\Delta t$, where $\Delta t = 1/[(1+\lambda)n(t)]$ and $n(t)$ is the
number of occupied vertices at time $t$. With probability
$p=1/(1+\lambda)$, the occupied vertex becomes vacant. With
complementary probability $1-p=\lambda/(1+\lambda)$, one of its $k_j$
neighbors is randomly selected and, if empty, it becomes occupied. If
the selected neighbor is already occupied nothing happens and the
simulation proceeds to the next step.

The standard numerical procedure to investigate the finite size
scaling at absorbing phase transitions is based on the determination
of the average of the order parameter (in this case the density of
active nodes), $\rho_s$, restricted only to surviving runs. Such a
technique is quite inefficient, because surviving configurations are
very rare at long times, and in order to get precise results an
exceedingly large number of realization of the process is needed. An
alternative strategy consists in constraining the system in a
quasi-stationary state and measuring its properties. In practice
this is implemented by replacing the absorbing state, every time the
system tries to visit it, with an active configuration randomly taken
from the history of the simulation~\cite{Mancebo2005}. For this task,
a list of $M$ active configurations is stored and constantly updated.
An update consists in randomly choosing a configuration in the list
and replacing it by the present active configuration with a
probability $p_{r}\Delta t$. After a relaxation time $t_r$, the QS
quantities are determined during an averaging time $t_a$. The QS
probability $\bar{P}_n$ that $n$ vertices are occupied is computed
during the averaging interval, each configuration with $n$ active
vertices contributing to the QS distribution with a probability
proportional to its lifespan, i.e. proportional to $1/n$. From the
particle distribution $\bar{P}_n$, the characteristic properties of
the QS state can be computed, such as the density of active nodes
\begin{equation}
 	\bar{\rho} = \frac{1}{N}\sum_{n\ge1}n\bar{P}_n,
\end{equation}
and the characteristic relaxation time~\cite{DickmanJPA,Ferreira_annealed}
\begin{equation}
	\tau = \frac{1}{\bar{P}_1}.
\end{equation}
The QS method, which had been
previously used to accurately determine the universality class of
several models with absorbing configurations
\cite{Mancebo2005,Dickman_sandpiles,Voronoi2008}, was recently applied
to investigate the critical properties of the CP in annealed SF
networks \cite{Ferreira_annealed}.

The FSS method for non-equilibrium absorbing phase transitions usually
assumes, at the transition, an asymptotic power-law dependence with the
system size of the order parameter, $\rho_s \sim N^{-\hat{\nu}}$, and of
the characteristic temporal scale, $\tau \sim N^{\hat{\alpha}}$. Values
of the control parameter above and below the critical point give rise
to positive and negative curvatures in plots of $\ln\rho_s$ versus
$\ln N$, respectively, for large $N$ \cite{marro1999npt}. The CP in annealed networks
has a different, anomalous, scaling
form~\cite{bogunaPRE2009,RomuPRL2008,NohPRE2009} which includes an
explicit dependence on the factor $g=\langle k^2\rangle/\langle k
\rangle^2$. In Ref.~\cite{Ferreira_annealed} the same form was
analytically shown to hold on annealed networks for the QS density
$\bar{\rho}$ and characteristic time $\tau$, namely
\begin{equation}
\label{eq:rhoscl}
\bar{\rho} \sim(gN)^{-1/2}, \quad
\tau \sim\left(\frac{N}{g}\right)^{1/2}.
\end{equation}
The presence of the factor $g$, which can be determined exactly for a
given degree distribution $P(k)$, brings an additional dependence on
$N$ for $\gamma<3$. Hence, the exponents predicted by HMF theory 
for the critical properties as a function of $N$ are \cite{RomuPRL2008}:
\begin{eqnarray}
 \hat{\nu} &=& \frac{1}{2}+\max\left(\frac{3-\gamma}{2\omega},0\right),	\label{eq:hmfexp1}
 \\ 
 \hat{\alpha}&=&\frac{1}{2}-\max\left(\frac{3-\gamma}{2\omega},
0\right).
\label{eq:hmfexponentpredictions}
\end{eqnarray}

\section{Results}
\label{sec:simu}

The QS simulations are performed on networks with sizes ranging from $N=10^4$ up
to $N=2\times 10^7$ and up to times $t_s=2\times 10^6$. The so-called structural
($\omega=2$) and natural ($\omega=\gamma-1$) hard cutoffs~\cite{mariancutofss}
are considered. The structural cutoff leads to the absence of degree-degree
correlations \cite{mariancutofss} in the random CM. In this case, the procedure
to build the network corresponds to the uncorrelated configuration model
(UCM)~\cite{Catanzaro05}. On the contrary, the ordered CM has, for $\gamma<3$,
disassortative degree-degree correlations, which are disregarded in the HMF
analysis~\cite{barratbook,dorogovtsevRMP08}. The determination of the QS
distributions starts after a relaxation time $t_r=10^6$. The dynamics is run for
several independent network realizations in order to average over many
topological configurations. Between $200$ and $500$ network samples were used
for each value of $N$. It is important to notice that the critical relaxation
time is very short when compared to the same quantity for regular lattices below
the critical dimension, due to the small world property of complex networks.

\subsection{Determination of the critical point}

In a quenched network the investigation of the transition is made
difficult by the presence of dynamical correlations, which shift the
position of the critical point. From the analytical point of view the
simplest way to take them into account is by means of a homogeneous pair
approximation, which yields~\cite{MunozPRL2010} $\lambda_c = \langle k
\rangle /(\langle k \rangle -1)$.

The criterion of null curvature, usually applied to find numerically
the critical point in regular lattices (see Sec.~\ref{sec:model}) was
used for the CP on quenched networks~\cite{RomuPRL2006,HongPRL2007}.
However, as shown for annealed networks~\cite{Ferreira_annealed}, such
method for estimating the critical point can be misleading if
anomalous scaling of the form~(\ref{eq:rhoscl}) occurs.
Additionally, dynamical analyses like spreading experiments
and density decay in time~\cite{marro1999npt} require absence of
finite size effects which are unavoidable in random networks with
the small world property~\cite{watts98,bogunaPRE2009}.
\begin{figure}[t]
 \centering
 \includegraphics[width=8.2cm]{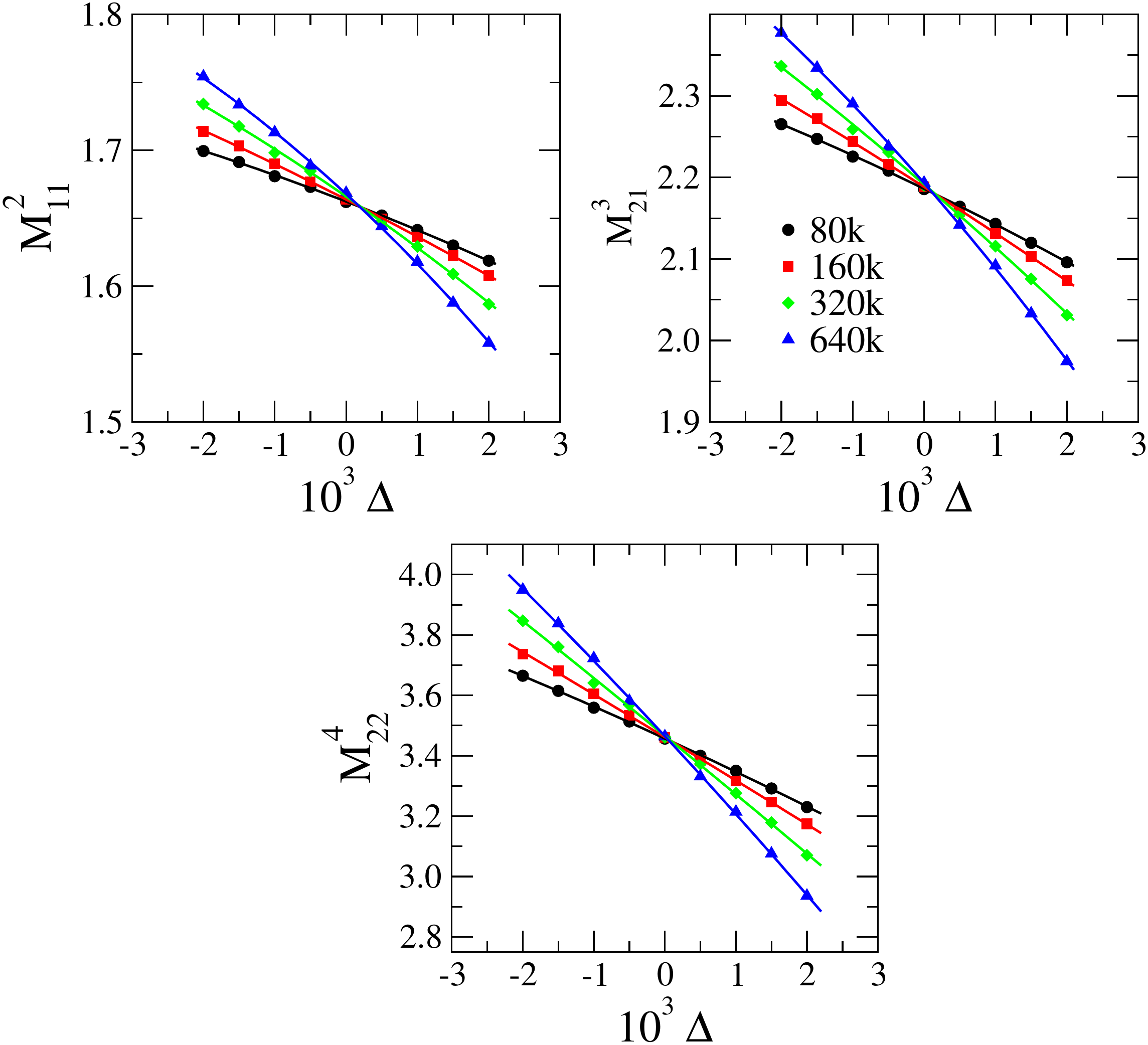}
 % momg275ann.pdf: 755x573 pixel, 72dpi, 26.63x20.21 cm, bb=0 0 755 573
\caption{(Color online) Moment ratios as function of the distance to
the critical point $\Delta=\lambda-\lambda_c$ for CP on annealed SF
networks with degree exponent $\gamma=2.75$ and cutoff exponent
$\omega=2$. Network sizes $N=8 \times 10^4$, $16 \times 10^4$, $32
\times 10^4$, and $64 \times 10^4$ are shown.}
 \label{fig:momann}
\end{figure}
For these reasons, we determine the position of the critical point by analyzing
the ratio between moments of the order parameter.
The well-known fourth order Binder reduced cumulant
$U_4=1-\langle \rho^4\rangle/3\langle \rho^2\rangle^2$
is a standard quantity to determine critical points in equilibrium
phase transitions on magnetic systems~\cite{landaubook}.
At criticality, this cumulant does not depend on
size, implying that curves $U_4(\lambda;N)$ against the control parameter
$\lambda$ (the temperature in magnetic systems) for different sizes all cross
at $\lambda=\lambda_c$.
More in general, moment ratios defined as 
\begin{equation}
M^{n}_{qs}=\frac{\langle\rho^n\rangle}{\langle\rho^q\rangle\langle \rho^s\rangle},~~~~q+s=n.
\end{equation}
are expected to be size-independent (and thus crossing each other) at
the transition. Universal and size-independent moment ratios were
studied for absorbing phase transitions in lattice models
\cite{Dickman_moments,Janssen2005,Sander2009}.
{The size independence of moment ratios in lattices systems 
results from the scaling invariance close to the critical point
(see e.g., Ref.~\cite{Henkel} for a recent review).}  
The order parameter in
the CP, the density of active vertices, is positive definite and thus,
in contrast with the magnetic counterpart for which all odd moments
are null due to symmetry, there is no restriction on the values of $q$
and $s$ that can be considered.
\begin{figure}[t]
 \centering
 \includegraphics[width=8cm]{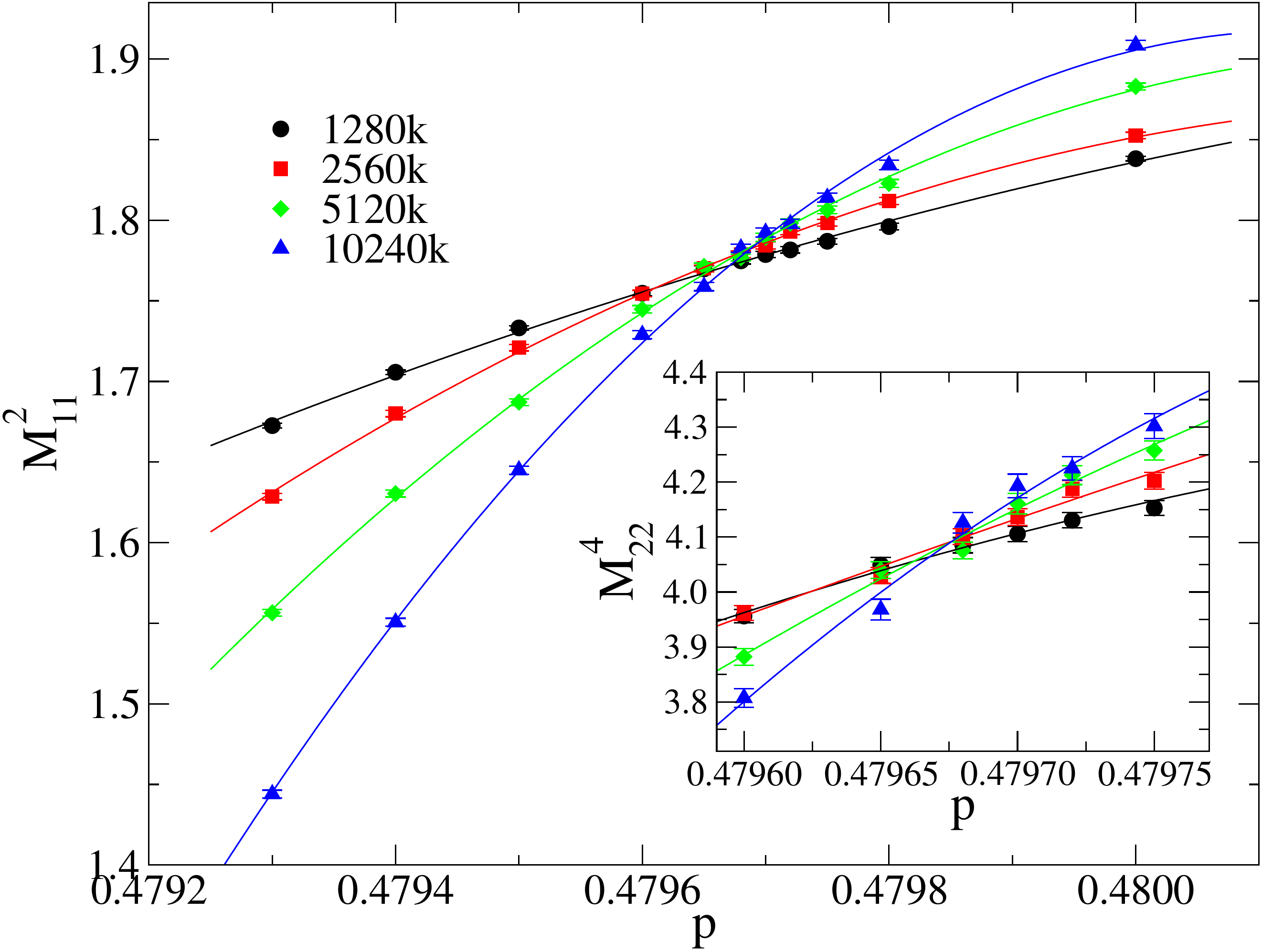}
 % momg275ann.pdf: 755x573 pixel, 72dpi, 26.63x20.21 cm, bb=0 0 755 573
 \caption{(Color online) Main plot: Moment ratios
$\langle\rho^2\rangle/\langle\rho\rangle^2$ as a function of the  annihilation
probability for CP on UCM networks with degree exponent $\gamma=2.75$
and minimum degree $k_0=6$. Symbols represent  stochastic
simulations, while solid lines are hyperbolic tangent regressions, drawn as
guides to the eyes. Inset: The corresponding fourth order moment
ratios $\langle\rho^4\rangle/\langle\rho^2\rangle^2$ around the
critical point. Network sizes $N=1.28 \times 10^6$, $2.56 \times 10^6$, $5.12 \times 10^6$, and $1.024 \times 10^7$ are shown.}
\label{fig:momg275}
\end{figure}

In order to probe the validity of this method for finding the critical
point in complex networks, we first determine moment ratios for the CP
in annealed SF networks, for which the critical point is exactly known
to be $\lambda_c=1$ \cite{bogunaPRE2009}. Figure~\ref{fig:momann} shows
the moment ratios $M^n_{qs}$ for CP on annealed networks with degree
exponent $\gamma=2.75$ and different sizes. As observed for
macroscopic quantities like density or characteristic time in annealed
networks~\cite{Ferreira_annealed}, moment ratios also have finite size
corrections; the crossing points converge to $\lambda=1$ strictly only
for $N\rightarrow\infty$. The convergence is fast, particularly for
the higher order moment ratios, which, however, are more susceptible
to large statistical fluctuations. For this reason we analyze moment
ratios up to fourth order. We find that the crossing points of moment
ratios tend to constant values, i.e. independent on the network
degree distribution (data not shown).  The second order moment ratio $M_{11}^2=1.667(3)$
is slightly larger than the value 1.660 obtained for CP on the
complete graph \cite{DickmanJPA}. The other moment ratios determined
are $M_{21}^3=2.190(4)$ and $M_{22}^4=3.452(3)$, respectively.
\begin{figure}[t]
	\centering
	\includegraphics[width=8.2cm]{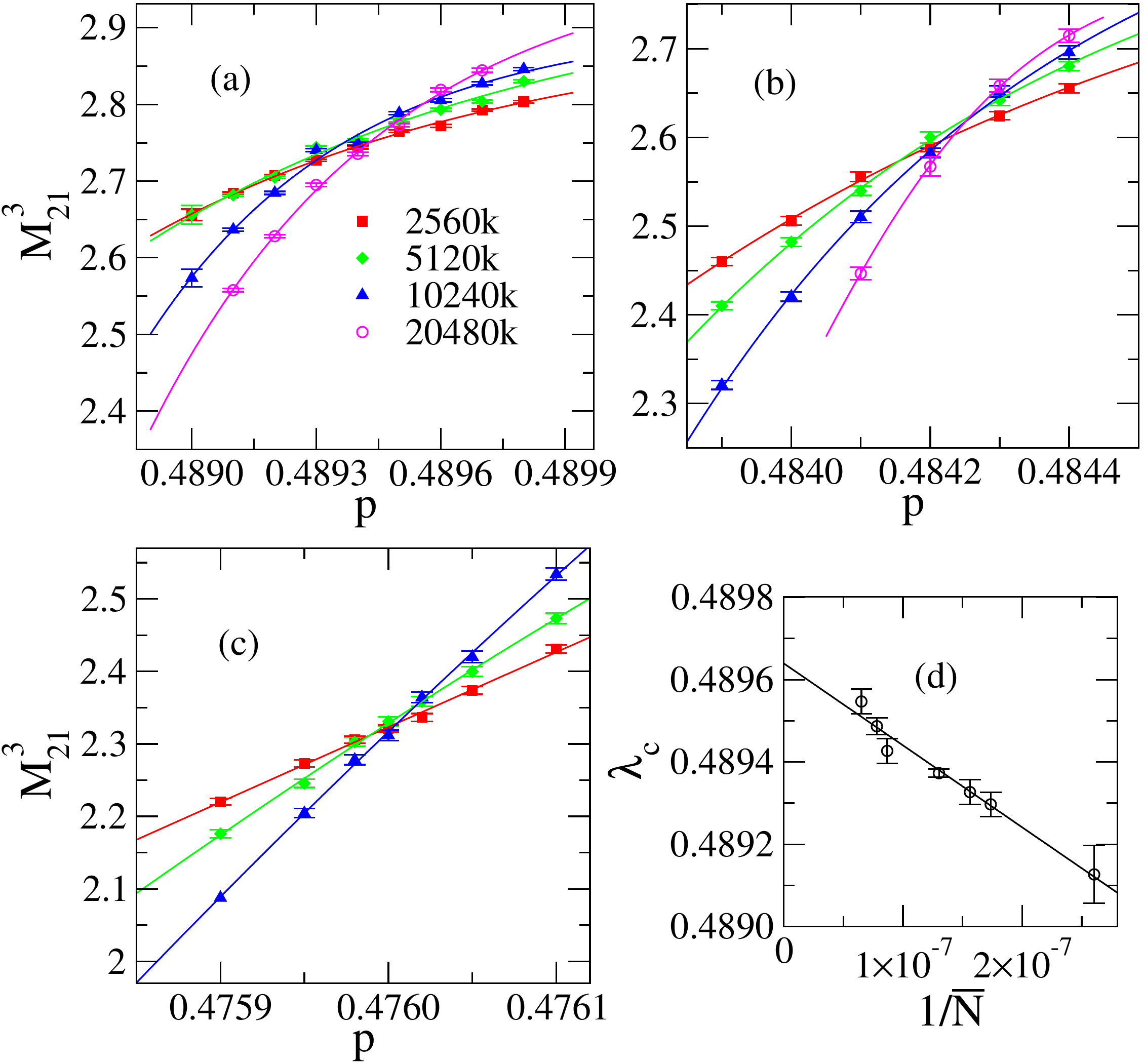}
	% momg275ann.pdf: 755x573 pixel, 72dpi, 26.63x20.21 cm, bb=0 0 755 573
	\caption{(Color online) Third order moment ratios for CP on UCM networks with
degree exponents: (a) $\gamma=2.25$, (b) $\gamma=2.50$, and (c)
$\gamma=3$. The minimum degree is $k_0=6$. Symbols represent
stochastic simulations, while solid lines are hyperbolic tangent
regressions, drawn as guides to the eyes. (d) Extrapolation used to
determine the critical point for $\gamma=2.25$. Network sizes $N=2.56
\times 10^6$, $5.12 \times 10^6$, and $1.024 \times 10^7$ are shown.}
 \label{fig:momgamas}
\end{figure}

In the case of quenched networks, the moment ratios for CP on UCM networks
($\omega=2$) with $\gamma=2.75$ and $k_0=6$ versus the annihilation probability
$p=1/(1+\lambda)$ are plotted in Fig.~\ref{fig:momg275}. Moment ratios are
determined for each network realization, and then averaged over the ensemble of
networks considered. The simulation data are very accurate, allowing to clearly
resolve the crossing points taking place between $0.47965$ and $0.47970$ for all
moment ratios analyzed (including the third order which is not shown.) As a
consequence the critical point position is estimated to be $p_c=0.47968(3)$ or,
equivalently, $\lambda_c=1.0847(1)$, where the number in parenthesis represent
the numerical uncertainty in the last digit. The critical moment ratios are
similarly estimated, yielding the values $M_{11}^2=1.77(2)$,
$M_{21}^3=2.42(2)$ and $M_{22}^4=4.10(9)$.

Third order moment ratios for different values of $\gamma$ are shown in
Fig.~\ref{fig:momgamas}(a)-(c). We observe that the higher the network
heterogeneity, the stronger the finite size effects; thus larger systems are
required for small $\gamma$ in order to determine the critical point with little
uncertainty. The critical points are determined by extrapolating the crossing
points between curves corresponding to sizes $N_a$ and $N_b$ against the inverse
of the mean size $1/\bar{N}=2/(N_a+N_b)$, assuming a functional dependence
$\lambda_c(\bar{N}) = \lambda_c + \mathrm{const}/\bar{N}$, as shown in Fig.
\ref{fig:momgamas}(d) for $\gamma=2.25$. The critical points and moment ratios
computed for different degree exponents are shown in Table \ref{tab:mom}.
{It is worth to stress that if we replace $\bar{N}^{-1}$ 
in the FSS by $\bar{N}^{-b}$, where $b$ is a fit parameter, the estimates
vary inside estimated error bars. For the most extreme case, $\gamma=2.25$,
we obtain $b=0.72$ and the critical point changes to $\lambda_c =
0.4897$.}
Our simulations show that the moment ratio analysis is
a very efficient method to determine the critical point of absorbing
phase transitions in complex networks. The moment ratios, in
opposition to annealed networks, depend now on the degree exponent through
an exponential form: As $\gamma$ increases, they approach the value
of the annealed case. The position of the critical point, on other
hand, is remarkably close to the predictions of pair approximation
calculations~\cite{MunozPRL2010} and it approaches the annealed value
$p_c=1/2$ for $\gamma\rightarrow 2$ when $\langle k \rangle\rightarrow\infty$.
\begin{table}[t]
\centering
 \begin{tabular}{cccccc}\hline\hline
 $\gamma$ & $p_c$& $p_c^{PA}$ & $M_{11}^2$&$M_{21}^3$ & $M_{22}^4$ \\\hline
  %2.15&?&?&?&? \\
  2.25&0.4896(1) & 0.48861 &~1.92(1)~&~2.77(3)~&~5.1(1)~\\
  2.50&0.48425(5)& 0.48386 &~1.85(2)~&~2.62(4)~&~4.7(1)~\\
  2.75&0.47968(3)& 0.47969 &~1.77(2)~&~2.42(4)~&~4.10(9)~\\
  3.00&0.47602(3)& 0.47628 &~1.72(2)~&~2.32(3)~&~3.81(7)~\\ 
  3.25&0.47303(3)& 0.47356 &~1.70(1)~&~2.25(2)~&~3.60(7)~\\ 
  ANN & 1/2      &1/2&~1.667(3)~ & ~2.190(4)~ & ~3.452(3)~ \\\hline\hline
 \end{tabular}
\caption{Critical points $p_c=1/(1+\lambda_c)$ and critical
moment ratios for CP on UCM networks with minimum degree $k_0=6$. 
The third column reports the predictions of the pair approximation
$p_c^{PA} = (\langle k \rangle-1) /(2\langle k \rangle -1)$.
The  values corresponding to annealed networks (ANN) are also included
for sake of comparison.}
\label{tab:mom}
\end{table}
\begin{figure}[hbt]
 \centering
%\subfigure[\label{fig:rhocritm6}]
{\includegraphics[width=7.5cm]{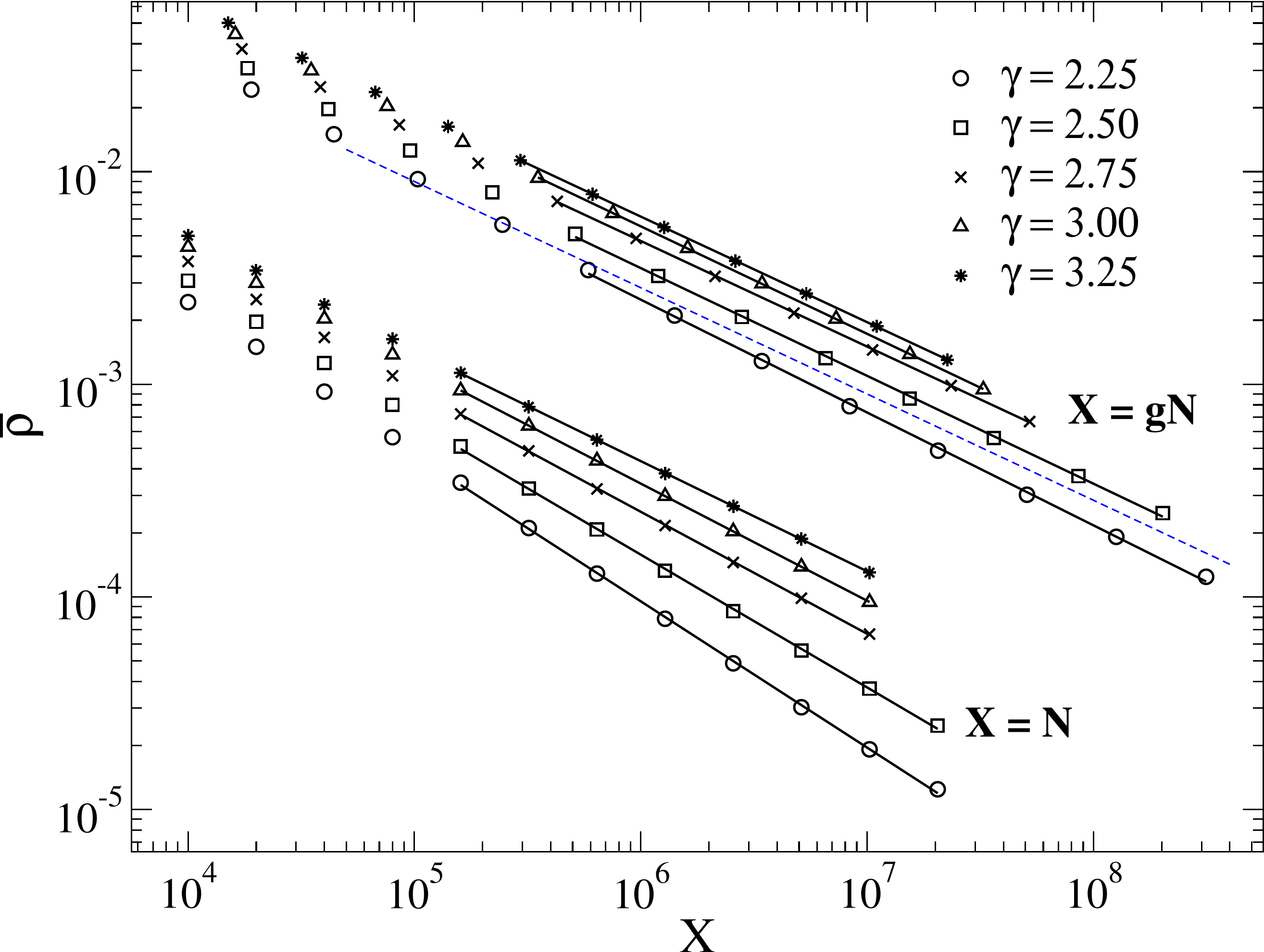}} 

~

%\subfigure[\label{fig:taucritm6}]
{\includegraphics[width=7.5cm]{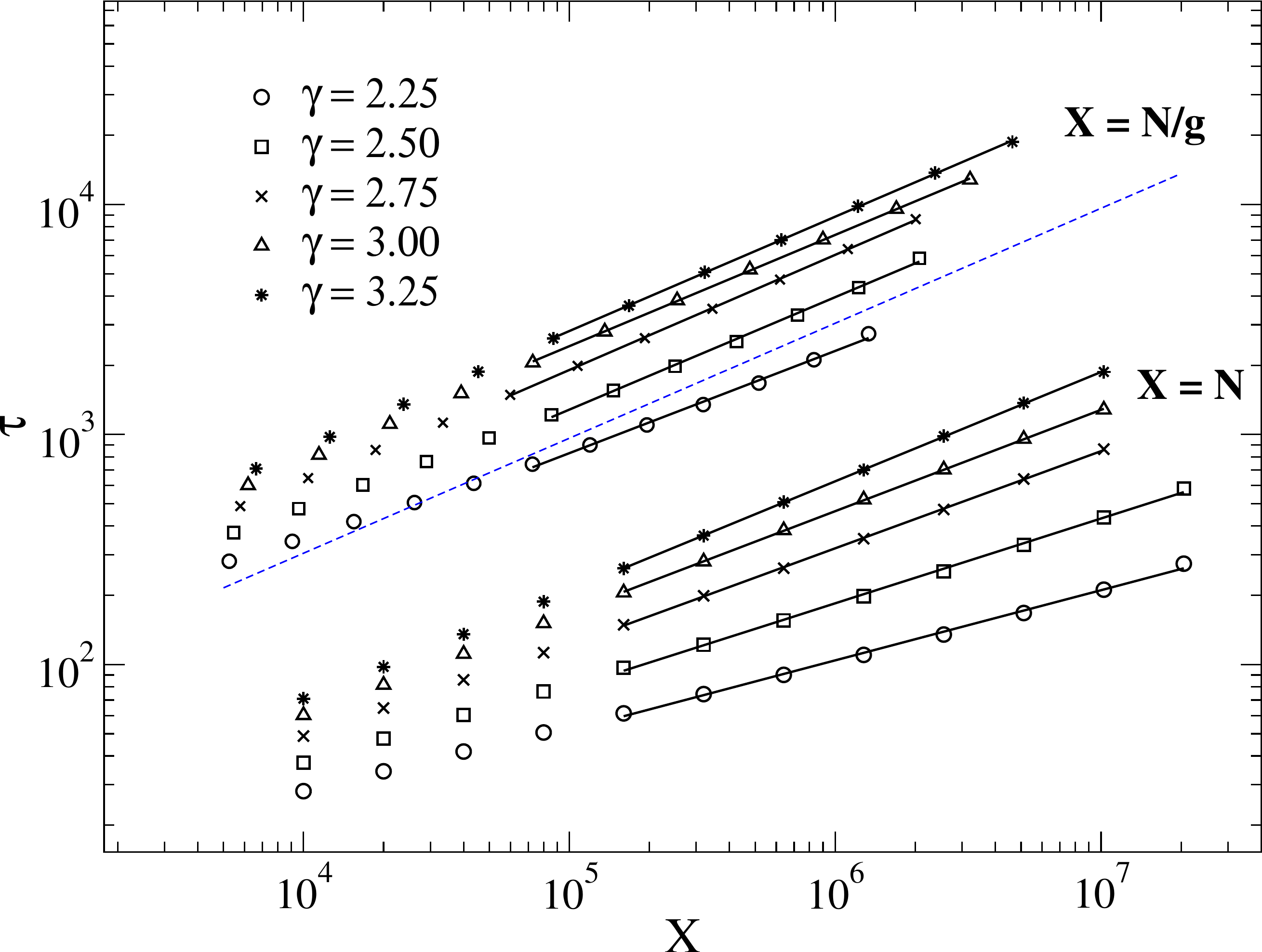}}
 % rhocrit_gamask06.pdf: 760x564 pixel, 72dpi, 26.81x19.90 cm, bb=0 0 760 564
 \caption{(Color online) Critical QS quantities (density in the top and characteristic time in
the bottom) for the UCM network with minimum degree $k_0=6$ and distinct degree
exponents. Symbols represent QS simulations and solid lines the respective power
law regressions. Dashed lines have slopes $\pm1/2$.}
\label{fig:qscrit}
\end{figure}

\subsection{Critical exponents}

\begin{figure}[t]
 \centering
 {\includegraphics[width=7.5cm]{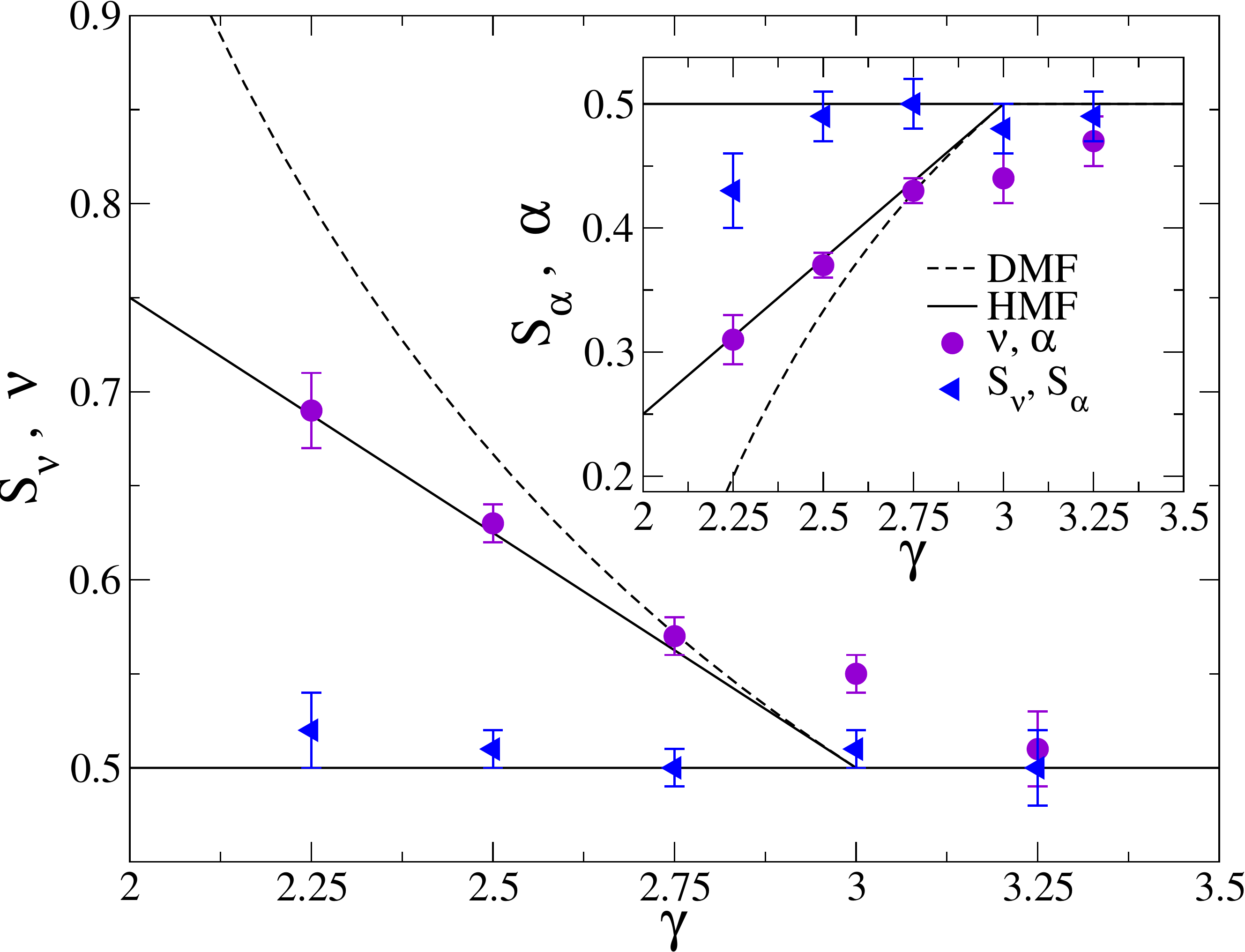}} ~ 
 % expovsgammako6.pdf: 738x576 pixel, 72dpi, 26.03x20.32 cm, bb=0 0 738 576
 \caption{(Color online) FSS exponents for CP on UCM networks with minimum
degree $k_0=6$, compared with the HMF and droplet mean field
(DMF)~\cite{HongPRL2007} theory predictions. Main panel shows the results of
power law regressions $\bar{\rho}\sim N^{-\hat{\nu}}$ and $\bar{\rho}\sim
(gN)^{-S_{\nu}}$ while the inset shows similar analysis for the characteristic
time $\tau$.}
 \label{fig:expk06}
\end{figure}

An accurate knowledge of the critical point allows to determine the
exponents characterizing the singular behavior at the transition and
also to probe the presence of corrections to the scaling. Assuming
that Eqs.~(\ref{eq:rhoscl}) hold also for quenched networks, we
perform a power law regression analysis in the plots of $\ln
\bar{\rho}$ vs $\ln N$ and $\ln \bar{\rho}$ vs $\ln gN$, as well as a
similar analysis for the characteristic time $\tau$. Both are shown in
Fig.~\ref{fig:qscrit}. 
{System sizes smaller than $N = 10^5$ are excluded from the regression to
avoid subleading corrections to scaling.}
The exponents obtained numerically as well as the HMF predictions in
Eqs.~\eqref{eq:hmfexp1} and \eqref{eq:hmfexponentpredictions} are
shown in Table~\ref{tab:expo} and Fig.~\ref{fig:expk06}.

For $\gamma=2.75$, $3$ and $3.25$, the
exponents $\hat{\nu}$ and $\hat{\alpha}$ estimated from a direct power
law regressions as a function of $N$ are quantitatively consistent
with those measured for annealed networks~\cite{Ferreira_annealed} but
they are larger and smaller, respectively, than the HMF values. This
is due to the fact that HMF exponents [Eq.~(\ref{eq:hmfexponentpredictions})]
are calculated by considering the asymptotic scaling
of $g$, $g \sim k_c^{3-\gamma}$. However, the
most relevant corrections to scaling
\begin{equation}
g \simeq \text{const} \times \left[1-\left(\frac{k_0}{k_c}\right)^{3-\gamma} +
2 \left(\frac{k_0}{k_c} \right)^{\gamma-2} \right] k_c^{3-\gamma}
\label{eq:correct}
\end{equation}
are nonnegligible even for the largest system sizes considered here.
Notice that for both $\gamma \to 2$ and $\gamma \to 3$ corrections
to scaling decay only logarithmically.
If instead we perform the regressions $\bar{\rho}\sim(gN)^{-S_\nu}$
and $\tau\sim(N/g)^{S_\alpha}$ we obtain  exponents very close to $0.5$, 
in full agreement with the HMF prediction $S_{HMF}=1/2$.
\begin{table*}[t]
 \begin{tabular}{cccccccccc}
 \hline\hline
 $\gamma$&$\nu$&$\nu_{HMF}$&$\alpha$&$\alpha_{HMF}$&
 $S_\nu$&$S_\alpha$ &$S_{HMF}$ \\ \hline
%2.25&~0.667(19)&~0.6875&~0.328(21)&~0.3125&~0.510(14) &~0.474(28)&~1/2 \\
%2.25&~0.67(2)&~0.6875&~0.33(2)&~0.3125&~0.51(1) &~0.47(3)&~1/2 \\
2.25&~0.69(2)&~0.6875&~0.31(2)&~0.3125&~0.52(2) &~0.43(3)&~1/2 \\
%2.50&~0.628(09)&~0.6250&~0.365(11)&~0.3750&~0.510(09) &~0.485(18)&~1/2 \\
2.50&~0.63(1)&~0.6250&~0.37(1)&~0.3750&~0.51(1) &~0.49(2)&~1/2 \\
%2.75&~0.572(10)&~0.5625&~0.425(13)&~0.4375&~0.495(09) &~0.503(15)&~1/2 \\
2.75&~0.57(1)&~0.5625&~0.43(1)&~0.4375&~0.50(1) &~0.50(2)&~1/2 \\
%3.00&~0.552(14)&~1/2   &~0.440(18)&~1/2   &~0.507(12) &~0.484(20)&~1/2 \\
3.00&~0.55(1)&~1/2   &~0.44(2)&~1/2   &~0.51(1) &~0.48(2)&~1/2 \\
%3.25&~0.514(18)&~1/2   &~0.472(15)&~1/2   &~0.492(17) &~0.493(16)&~1/2 \\ 
3.25&~0.51(2)&~1/2   &~0.47(2)&~1/2   &~0.49(2) &~0.49(2)&~1/2 \\ 
 \hline\hline
 \end{tabular}
\caption{FSS exponents for the CP on UCM network with $k_0=6$. HMF
exponents, Eq.~\eqref{eq:hmfexponentpredictions}, are also shown for
comparison. The number in parenthesis represents the error in the last
digit given by  the standard deviation of the  exponents fitted 
for the two curves  that are closest
to the critical point, being one below and other above
$\lambda_c$.}
\label{tab:expo}
\end{table*}

For $\gamma=2.50$, the corrections to the scaling of factor $g$ vanish
as $N^{-1/4}$, the fastest decay in the analyzed $\gamma$ range. Both
characteristic time and density exponents agree with HMF theory if the
factor $g$ is explicitly included. If $g$ is not included, the exponents
still agree with the HFM predictions, due to the fast decay of the
correction to scaling in $g$.
{For $\gamma=2.25$, the exponents obtained with a direct power law
regression exhibit an excellent agreement with HMF theory,
with the exception of the scaling of $\tau$ vs $N/g$, probably due to
finite size effects induced by preasymptotic corrections for the
mean degree $\langle k \rangle$.}

\subsection{The natural cutoff}

\begin{figure}[tb]
 \centering
 \includegraphics[width=8.0cm]{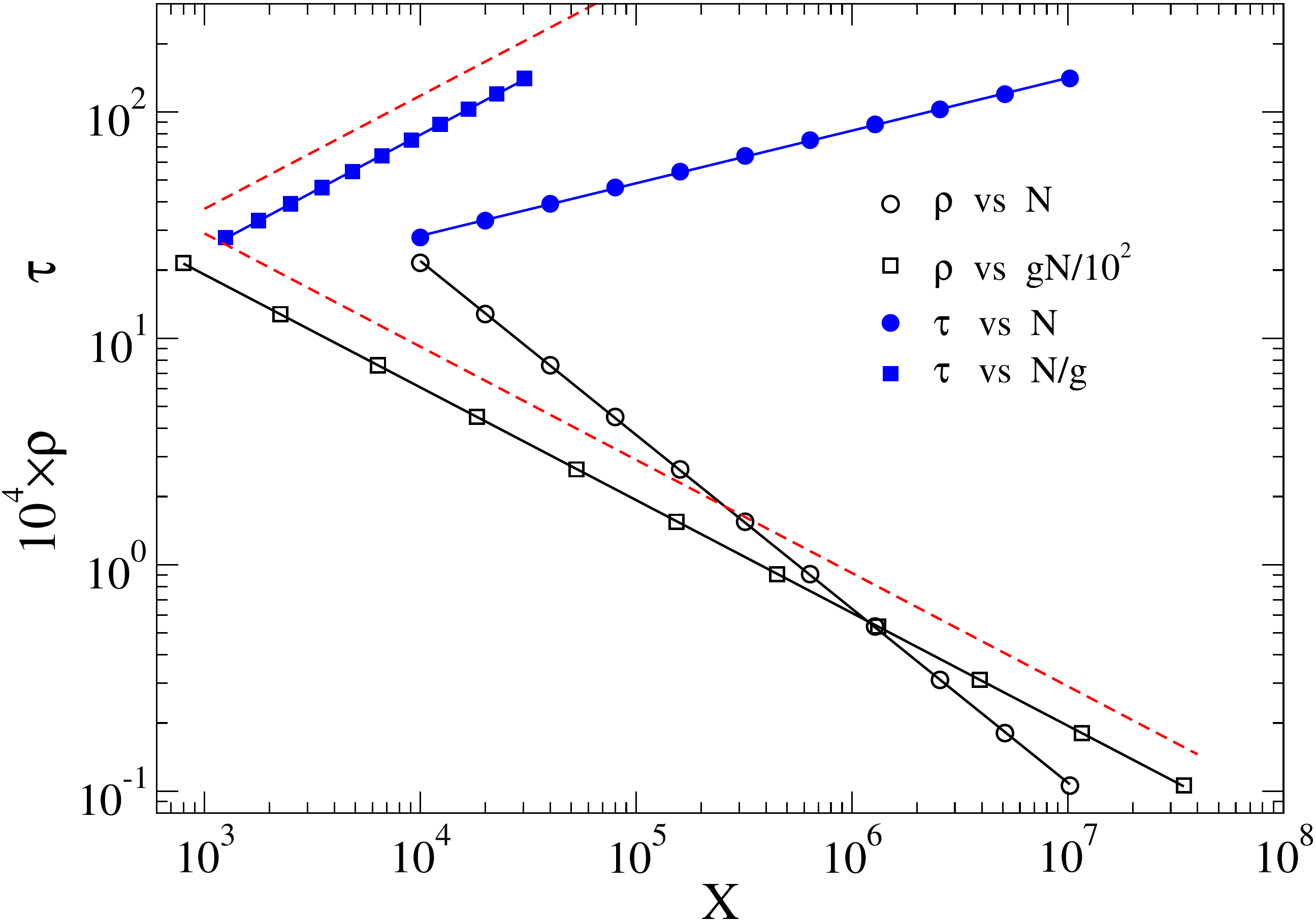}
 % qscrit_g225nat.pdf: 746x523 pixel, 72dpi, 26.32x18.45 cm, bb=0 0 746 523
 \caption{(Color online) Critical QS quantities for OCM networks with degree exponent $\gamma =
2.25$, cutoff exponent $\omega=\gamma-1$, and minimum degree $k_0=6$. Symbols
represent QS simulations, solid lines are power regressions and dashed lines power 
laws with exponents $\pm1/2$.}
 \label{fig:qsnat}
\end{figure}

The structural cutoff $\omega=2$ has been used to investigate dynamical process
\cite{RomuPRL2010,RomuPRL2006,HongPRL2007} in SF networks because it
prevents degree-degree correlations \cite{mariancutofss} and makes the HMF
approach analytically feasible. However, the slow growth of $k_c$ with $N$
implies large corrections to the asymptotic
scaling~\cite{Ferreira_annealed,bogunaPRE2009} and this makes in turn
the numerical analysis hard in highly heterogeneous quenched substrates.
For this reason we additionally simulate the CP on the ordered
configuration model (OCM) described in Sec.~\ref{sec:model} with a
hard cutoff $k_c=N^{1/(\gamma-1)}$. For this cutoff,
HMF predicts the critical exponents to be $\hat{\nu} =
\max[1/(\gamma-1),1/2]$ and $\hat{\alpha} = \max[(\gamma-2)/(\gamma-1),1/2]$.
We present only the results for highly heterogeneous case
$\gamma=2.25$ with $k_0=6$, since the conclusions hold also for other values
of $\gamma$.
The crossing points in the moment ratio analysis converge faster to the
asymptotic value,  when compared with the UCM.
We find a critical point $p_c=0.4908(1)$ and moment ratios
$M_{11}^2=1.714(4)$, $M_{21}^3=2.29(1)$, and $M_{22}^4 = 3.78(3)$,
which are closer to the HMF predictions than those shown in Table \ref{tab:mom}
for  UCM networks.
As can be seen in Fig.~\ref{fig:qsnat},
the QS quantities exhibit very good scaling properties even for highly
heterogeneous networks.
The critical exponents obtained by direct power law regressions
are $\hat{\nu}=0.774(5)$ and $\hat{\alpha}=0.227(5)$.
They differ from the HMF values $0.8$ and $0.2$, but also in this case the
discrepancy is due to the preasymptotic scaling of $g$. Indeed the
regressions including the factor $g$
yield $S_\nu=0.498(4)$ and $S_\alpha=0.502(5)$, in remarkable agreement
with HMF.
This excellent agreement is rather  surprising since the HMF exponents are
obtained under the hypothesis of no degree-degree
correlations~\cite{RomuPRL2008,Ferreira_annealed}, while OCM networks are 
disassortative~\cite{mariancutofss}.
The present findings suggest that this kind of correlations does not strongly
affect the behavior of the CP. 
{Further theoretical investigation is needed to clarify why topological
correlations play such a minor role in this problem.}

\section{Conclusions}

In this paper we have investigated the behavior of the
contact process on quenched networks with power-law distributed degrees.
By performing quasi-stationary simulations we have probed the
system behavior in the critical region, obtaining very precise
numerical results for both the position of the transition between
absorbing and active phases and the associated critical exponents.
We have taken advantage of the recent progress in the analytical
understanding of the anomalous finite size scaling behavior on
annealed networks.

It turns out that the quenched structure of the substrate
has little effect on the contact process dynamics.
Both the qualitative form of the scaling behavior and the
quantitative value of the exponents are the same for quenched
and annealed networks. The only, minor, modifications affect
the position of the critical point, which is shifted in
quenched networks due to dynamical correlations, as well as
the values of the moment ratios at criticality, which depend
on $\gamma$ on quenched networks, while are universal on annealed
ones.

The conclusion that can be drawn is that heterogeneous mean-field
theory correctly describes (within our error estimates) the critical
behavior of the contact process on quenched networks. This conclusion
corrects the apparent failure of HMF predictions for CP on quenched
networks that was reported some years ago in Ref.~\cite{RomuPRL2006}
%The conclusion that can be drawn is that heterogeneous mean-field
%theory correctly describes the critical behavior of the contact process
%on quenched networks. This conclusion ends a debate that
%started some years ago when the apparent failure
%of HMF predictions for CP on quenched networks was reported~\cite{RomuPRL2006}.
With the benefit of hindsight, it is clear now that the puzzling
discrepancy found in~\cite{RomuPRL2006} between HMF and numerical
results on quenched networks was not due to subtle effects of the
quenched topology, rather to the nontrivial anomalous nature
of finite size effects on annealed networks.

\begin{acknowledgments}
  This work was partially supported by the Brazilian agencies CNPq and
  FAPEMIG.  S.C.F thanks the kind hospitality at the Departament de
  F\'{\i}sica i Enginyeria Nuclear/UPC.  R.P.-S.  acknowledges
  financial support from the Spanish MEC, under project
  FIS2010-21781-C02-01; the Junta de Andaluc\'{\i}a, under project
  No. P09-FQM4682; and additional support through ICREA Academia,
  funded by the Generalitat de Catalunya.
 
\end{acknowledgments}

%\bibliography{cpquench}

\end{document}